\documentclass[aps,pra,reprint,a4paper,balancelastpage]{revtex4-1}

\pdfoutput=1

\usepackage[UKenglish]{babel}
\usepackage[UKenglish,cleanlook]{isodate}
\usepackage{amsmath}
\usepackage{amssymb}
\usepackage[retainorgcmds]{IEEEtrantools}
\usepackage{dsfont}
\usepackage{tikz}
\usepackage[colorlinks,linkcolor=blue,citecolor=blue,urlcolor=blue]{hyperref}
\usepackage{quant_defs}

\def\algor{Algorithmica}%
\def\natcomm{Nat.\ Commun.}%
\def\njp{New J.\ Phys.}%
\def\pla{Phys.\ Lett.\ A}%
\def\prsa{Proc.\ R.\ Soc.\ A}%
\def\quantic{Quantum Inf.\ Comput.}%

\usetikzlibrary{shapes.misc,shapes.symbols,shapes.geometric,arrows,
  plotmarks,decorations.pathmorphing}
\tikzset{gridlines/.style={very thin,color=gray!50},
  every picture/.append style={scale=0.7}}

\definecolor{darkgreen}{rgb}{0,0.5,0}

\newcommand{\vect}[1]{\boldsymbol{#1}}

\newcommand{\rA}{\mathrm{A}}
\newcommand{\rB}{\mathrm{B}}

\newcommand{\rE}{\mathrm{E}}
\newcommand{\rBE}{\mathrm{BE}}
\newcommand{\rABE}{\mathrm{ABE}}
\newcommand{\rEE}{\mathrm{EE}}
\newcommand{\hilb}[0]{\mathcal{H}}
\newcommand{\etal}[0]{\textit{et al.}}

\newcommand{\Hc}[0]{H} 
\newcommand{\Hq}[0]{H} 

\newcommand{\0}[0]{^{\vphantom{\prime}}}
\newcommand{\1}[0]{^{\prime}}

\newcommand{\work}{article}

\begin{document}

\title{Tight asymptotic key rate for the BB84 protocol \\ with local
  randomisation and device imprecisions}
\author{Erik Woodhead}
\date{4 August 2014}
\email{Erik.Woodhead@ulb.ac.be}
\affiliation{Laboratoire d'Information Quantique, CP~225, Universit\'e{}
  libre de Bruxelles, av.\ F.~D.~Roosevelt 50, 1050 Bruxelles, Belgium}

\begin{abstract}
  Local randomisation is a preprocessing procedure in which one of the
  legitimate parties of a quantum key distribution (QKD) scheme adds noise to
  their version of the key and was found by Kraus \etal{}
  [\href{http://dx.doi.org/10.1103/PhysRevLett.95.080501}{%
    \prl\ \textbf{95},\ 080501 (2005)}] to improve the security of certain
  QKD protocols. In this \work{}, the improvement yielded by local
  randomisation is derived for an imperfect implementation of the BB84 QKD
  protocol, in which the source emits four given but arbitrary pure states
  and the detector performs arbitrarily-aligned measurements. Specifically,
  this is achieved by modifying an approach to analysing the security of
  imperfect variants of the BB84 protocol against collective attacks,
  introduced in [\href{http://dx.doi.org/10.1103/PhysRevA.88.012331}{%
    \pra\ \textbf{88},\ 012331 (2013)}], to include the additional
  preprocessing step. The previously known improvement to the threshold
  channel noise, from 11\% to 12.41\%, is recovered in the special case of an
  ideal BB84 implementation and becomes more pronounced in the case of a
  nonideal source. Finally, the bound derived for the asymptotic key rate,
  both with and without local randomisation, is shown to be tight with the
  particular source characterisation used. This is demonstrated by the
  explicit construction of a family of source states and optimal attacks for
  which the key-rate bound is attained with equality.
\end{abstract}


\maketitle

\section{Introduction}

Quantum key distribution (QKD) \cite{ref:bb1984} was proposed three decades
ago as a potentially feasible way to generate and distribute cryptographic
keys in a secure manner, based on limitations inherent to quantum
physics. The possibility of secure QKD intuitively follows from the monogamy
of entanglement of quantum states or the impossibility of perfect state
cloning \cite{ref:wz1982,ref:d1982}, depending on the implementation.

The main theoretical problem in QKD consists in determining how many key bits
can be extracted securely from a given protocol, i.e., determining a bound on
the \emph{key rate}, particularly under realistic conditions such as the
presence of channel noise and imperfectly precise sources and detectors. This
is an incompletely solved problem even for the original protocol proposed by
Bennett and Brassard \cite{ref:bb1984}, now commonly called the BB84
protocol. For instance, in the case of an otherwise ideal implementation
suffering from channel noise, and if the key is extracted using one-way
postprocessing, the BB84 protocol has been shown to be secure asymptotically
if the error rate is less than the Shor-Preskill error rate of 11\%
\cite{ref:sp2000}, but has only been proved \emph{insecure} if the error rate
exceeds around 14.64\% \cite{ref:fg1997}. Kraus \etal{} found that the lower
threshold error rate could be increased to around 12.41\% by local
randomisation, in which one of the legitimate parties randomly flips a
fraction of their raw key bits as a preprocessing step
\cite{ref:kgr2005,ref:rgk2005}; Smith \etal{} subsequently increased this
threshold to 12.92\% using block codes \cite{ref:srs2008}. If two-way
postprocessing is used, the threshold error rate is known to lie between 20\%
and 25\% \cite{ref:c2002,ref:ba2007}. More recently, local randomisation has
been investigated for the BB84 protocol in the case of finite statistics
\cite{ref:mk2013}.

Parallel to this, substantial effort has gone toward adapting security proofs
for the BB84 protocol to account for device imprecisions, as described for
instance in Refs.~\cite{ref:kp2003,ref:gl2004,ref:k2009,ref:mls2010,
  ref:ly2011,ref:tl2012,ref:w2013}, where the goal is to derive a key-rate
bound in a setting where either the source or measurement basis states or
both are allowed to deviate from the $\sz$ and $\sx$ eigenstates ideally
required by the BB84 protocol. Typically, the result is a generalisation of
the Shor-Preskill key rate depending on additional parameters characterising
the quantum devices, and the improvement obtainable with preprocessing is not
investigated.

In this \work{}, an approach to accounting for source imprecisions for the
BB84 protocol, described in Ref.~\cite{ref:w2013}, is extended to include the
local randomisation preprocessing procedure described in
\cite{ref:kgr2005,ref:rgk2005}. The approach is based on the security
framework by Devetak and Winter \cite{ref:dw2005} and can be used to derive
bounds on the asymptotic key rate secure against an adversary restricted to
collective attacks \cite{ref:bb2002}. Conceptually, it separates into two steps:
first, a lower bound on the Devetak-Winter key rate is derived in terms of an
adversary's ability to distinguish their marginals of the $z$-basis states,
as measured by the fidelity; second, the fidelity itself is lower bounded in
terms of the observed error rate (such a bound can be viewed as quantifying
the measurement-disturbance tradeoff or the limits on state cloning imposed
by quantum physics). The incorporation of local randomisation presented here
concerns only the first of these steps and is achieved by a straightforward
generalisation of the intermediate bound on the key rate in terms of the
fidelity derived in \cite{ref:w2013}. The approach can therefore
automatically apply to any BB84-like protocol for which a fidelity bound of
the type derived in \cite{ref:w2013} is known or can be derived.

Following this adaptation, particular attention is given to a nonideal BB84
implementation in which one party (Alice) has a source transmitting four
characterised but arbitrary pure states and the second party (Bob)'s
measurements are uncharacterised. The relative improvement obtained with the
additional preprocessing is found to become more significant in the case of
an imperfect source. Finally, the bound obtained on the Devetak-Winter rate,
both with and without the additional preprocessing, is shown to be tight;
this is demonstrated by the explicit construction of a family of sources and
optimal collective attacks for which the Devetak-Winter rate is attained with
equality.

As mentioned above, the security results derived here assume an adversary
restricted to collective attacks. In Appendix~\ref{sec:virtual_ent}, we
outline how the (prepare-and-measure) BB84 protocol considered here could be
recast in an equivalent entanglement-based form, for which security against
collective attacks is already known to imply security against general attacks
under the assumption of a dimension bound on the Hilbert space.

\section{General scenario and method\label{sec:scenario}}

The following generic setting is considered: one party (Alice) possesses a
source capable of emitting one of a number of quantum states and transmitting
them over an untrusted quantum channel to a second party (Bob), who performs
measurements on them. Two of the states, which we will call the ``$z$-basis''
states (even if they are not orthogonal) and denote by $\ket{\alpha}$ and
$\ket{\alpha'}$, are intended for key generation. These should be selected
(equiprobably between them) by Alice and transmitted to Bob the majority of
the time. The remaining states are intended for the purpose of characterising
the quantum channel and testing for the presence of an eavesdropper. One of
Bob's possible measurements (a ``$z$-basis'' measurement) should be used a
majority of the time and should ideally be calibrated in such a way that the
measurement outcome is maximally correlated with Alice's choice of $z$-basis
state. After all the quantum states have been transmitted, Alice and Bob
publicly reveal in which cases they each used the $z$-basis states and
measurement. A subset of the results are sacrificed in order to estimate the
$z$-basis error rate, which we denote by $\delta_{z}$, after which they are
discarded. The cases where at least one of Alice or Bob did not use the $z$
basis are used to estimate one or more parameters, which we collectively
denote $\vect{\omega}$, depending on the specific details of the
protocol. Alice then flips a fraction $\eta$ of her $z$-basis key bits,
publicly revealing $\eta$ but not which bits she flips. Finally, Alice and
Bob extract a key using one-way error correction and privacy amplification,
as usual.

We assume that an adversary attacks the quantum channel unitarily,
individually, identically, in which case they acquire partial traces
$\rho\0_{\rE} = \Tr_{\rB}[\proj{\alpha}]$ or $\rho\1_{\rE} =
\Tr_{\rB}[\proj{\alpha'}]$ of the $z$-basis states transmitted by Alice, and
that they are allowed to delay their measurements indefinitely. In this case,
a lower bound on the extractable secret key rate by one-way postprocessing is
given by the Devetak-Winter rate \cite{ref:dw2005}, which we express as
\begin{equation}
  \label{eq:dw_rate}
  r = \Hq(Z \mid \rE) - \Hc(Z \mid Z') \,.
\end{equation}
In \eqref{eq:dw_rate}, $\Hc(Z \mid Z')$ denotes the classical (Shannon)
entropy of Alice's version of the key conditioned on Bob's $z$-basis
measurement outcome and quantifies the key lost by error correction. In the
typical case of symmetric errors, $\Hq(Z \mid Z') = h(\tilde{\delta}_{z})$,
where $\tilde{\delta}_{z} = (1 - \eta) \delta_{z} + \eta (1 - \delta_{z})$ is
the error rate between Alice and Bob's $z$-basis results (after Alice has
flipped a fraction $\eta$ of her bits), $h(p) = - p \log(p) - (1 - p) \log(1
- p)$ is the binary entropy, and throughout this \work{}, $\log$ is the
base-2 logarithm.

$\Hq(Z \mid \rE)$ denotes the von Neumann entropy of Alice's key bits
conditioned on the adversary's quantum side information, formally evaluated
on the classical-quantum state
\begin{IEEEeqnarray}{rCl}
  \label{eq:cq_state}
  \tau_{Z\rE}
  &=& \tfrac{1}{2} \bigro{(1 - \eta) \proj{0}_{Z} + \eta \proj{1}_{Z}}
  \otimes \rho\0_{\rE} \IEEEnonumber \\
  &&+\> \tfrac{1}{2} \bigro{\eta \proj{0}_{Z} + (1 - \eta) \proj{1}_{Z}}
  \otimes \rho\1_{\rE} \,.
\end{IEEEeqnarray}
In this framework, the main objective is to obtain a lower bound on $\Hq(Z
\mid \rE)$ in terms of the parameters $\vect{\omega}$ observed by Alice and
Bob. Following the approach in \cite{ref:w2013}, this can be separated into
two steps. We first derive a lower bound on $\Hq(Z \mid \rE)$ in terms of the
fidelity $F(\rho\0_{\rE}, \rho\1_{\rE}) = \trnorm{\sqrt{\rho\0_{\rE}}
  \sqrt{\rho\1_{\rE}}}$ of the adversary's marginal $z$-basis states. This
can then be complemented by a suitable bound $F(\rho\0_{\rE}, \rho\1_{\rE})
\geq F(\vect{\omega})$ on the fidelity itself. Such a bound will generally
depend on the details of the specific protocol being considered and
assumptions about the source states and/or Bob's measurements. For example,
for an ideal BB84 implementation with no local randomisation, the
Shor-Preskill key rate is recovered by combining $\Hq(Z \mid \rE) \geq 1 -
h\bigro{\tfrac{1}{2} + \tfrac{1}{2} F(\rho\0_{\rE}, \rho\1_{\rE})}$ with
$F(\rho\0_{\rE}, \rho\1_{\rE}) \geq \abs{1 - 2 \delta_{x}}$, where
$\delta_{x}$ is the $x$-basis error rate \cite{ref:w2013}.

In order to derive a lower bound on $\Hq(Z \mid \rE)$, we first reexpress the
classical-quantum state \eqref{eq:cq_state} as
\begin{equation}
  \label{eq:cq_mixed}
  \tau_{Z\rE} = \tfrac{1}{2} \proj{0}_{Z} \otimes \tilde{\rho}\0_{\rE}
  + \tfrac{1}{2} \proj{1}_{Z} \otimes \tilde{\rho}\1_{\rE} \,,
\end{equation}
where $\tilde{\rho}\0_{\rE} = \Tr_{\rB}[\tilde{\rho}]$ and $\tilde{\rho}\1_{\rE}
= \Tr_{\rB}[\tilde{\rho}\1]$, and we have set
\begin{IEEEeqnarray}{rCl}
  \tilde{\rho}\0 &=& (1 - \eta) \rho\0 + \eta \rho\1 \,, \\
  \tilde{\rho}\1 &=& \eta \rho\0 + (1 - \eta) \rho\1 \,,
\end{IEEEeqnarray}
where $\rho\0 = \proj{\alpha\0}$ and $\rho\1 = \proj{\alpha\1}$ are the two
$z$-basis states. The conditional von Neumann entropy, evaluated directly on
the classical-quantum state \eqref{eq:cq_mixed}, simplifies to
\begin{equation}
  \label{eq:H_ZE_rhos}
  \Hq(Z \mid \rE) = 1 + \tfrac{1}{2} \bigro{S(\tilde{\rho}\0_{\rE})
    + S(\tilde{\rho}\1_{\rE})}
  - S\bigro{\tfrac{1}{2} (\rho\0_{\rE} + \rho\1_{\rE})} \,,
\end{equation}
with $S(\rho) = - \Tr[\rho \log(\rho)]$. Following the approach in
\cite{ref:w2013}, we use that $\Hq(Z \mid \rE) \geq \Hq(Z \mid \rEE')$ for
any extension $\tau_{Z\rEE'}$ of the state \eqref{eq:cq_mixed} to a larger
Hilbert space in order to replace $\rho\0_{\rE}$ and $\rho\1_{\rE}$ with
purifications $\ket{\psi}$ and $\ket{\psi'}$ chosen such that
$F(\rho\0_{\rE}, \rho\1_{\rE}) = \braket{\psi}{\psi'}$. With this
substitution,
\begin{IEEEeqnarray}{rCl}
   \label{eq:H_ZE_S_pure}
  \Hq(Z \mid \rE) &\geq& 1
  + \tfrac{1}{2} S \bigro{(1 - \eta) \proj{\psi}
    + \eta \proj{\psi'}} \IEEEnonumber \\
  &&+\> \tfrac{1}{2} S \bigro{\eta \proj{\psi}
    + (1 - \eta) \proj{\psi'}} \IEEEnonumber \\
  &&-\> S \bigro{\tfrac{1}{2} (\proj{\psi} + \proj{\psi'})} \,.
\end{IEEEeqnarray}
The eigenvalues of the operator $(1 - \eta) \proj{\psi} + \eta \proj{\psi'}$
are easily found to be $\tfrac{1}{2} \pm \tfrac{1}{2} \sqrt{1 - 4 \eta (1 -
  \eta) (1 - \abs{\braket{\psi}{\psi'}}^{2})}$. Consequently,
\begin{multline}
  \label{eq:H_ZE_noise}
  \Hq(Z \mid \rE) \geq 1
  - h \bigro{\tfrac{1}{2} + \tfrac{1}{2} F(\rho\0_{\rE},
    \rho\1_{\rE})} \\
  +\> h \Bigro{\tfrac{1}{2}
    + \tfrac{1}{2} \sqrt{1 - 4 \eta (1 - \eta)
      \bigro{1 - F(\rho\0_{\rE}, \rho\1_{\rE})^{2}}}} \,.
\end{multline}
In Appendix~\ref{sec:convexity}, the right-hand side of \eqref{eq:H_ZE_noise}
is shown to be an increasing function of the fidelity. Given a lower bound
$F(\rho\0_{\rE}, \rho\1_{\rE}) \geq F(\vect{\omega})$ on the fidelity, then,
we obtain the analytic lower bound
\begin{IEEEeqnarray}{rCl}
  \label{eq:r_fidel_bound}
  r &\geq& 1 + h \Bigro{\tfrac{1}{2}
    + \tfrac{1}{2} \sqrt{1 - 4 \eta (1 - \eta)
      \bigro{1 - F(\vect{\omega})^{2}}}} \IEEEnonumber \\
  &&-\> h \bigro{\tfrac{1}{2} + \tfrac{1}{2} F(\vect{\omega})}
  - h\bigro{(1 - \eta) \delta_{z} + \eta (1 - \delta_{z})}
\end{IEEEeqnarray}
for the key rate.

The best result for the key rate is obtained by maximising the right-hand
side of \eqref{eq:r_fidel_bound} over $\eta$, which, if necessary, is readily
done numerically. Typically, as the channel noise approaches the maximal
threshold, the optimal fraction $\eta$ approaches $1/2$. In this regime, the
behaviour of the key-rate bound can be studied by substituting $\eta = (1 -
\varepsilon) / 2$ and expanding the resulting expression in powers of
$\varepsilon$. The result, to the first non-trivial order in $\varepsilon$,
is
\begin{equation}
  r \gtrsim \biggro{
    - \frac{1 - F(\vect{\omega})^{2}}{4 F(\vect{\omega})}
    \log \biggro{\frac{1 + F(\vect{\omega})}{1 - F(\vect{\omega})}}
    + \frac{(1 - 2 \delta_{z})^{2}}{2 \ln(2)}} \varepsilon^{2} \,,
\end{equation}
where $\ln$ is the natural logarithm. Consequently, threshold error rates can
be obtained by identifying corresponding roots of the expression
\begin{equation}
  \label{eq:thresh_root}
  \bigro{1 - F(\vect{\omega})^{2}} \ln \biggro{
    \frac{1 + F(\vect{\omega})}{1 - F(\vect{\omega})}}
  - 2 F(\vect{\omega}) (1 - 2 \delta_{z})^{2} \,.
\end{equation}

\section{Result for nonideal BB84\label{sec:pmbb84}}

The technique described in the preceding section can, in principle, be
applied to any BB84-like protocol for which a bound of the form
$F(\rho\0_{\rE}, \rho\1_{\rE}) \geq F(\vect{\omega})$ for the fidelity is
known or can be derived. In this section, we illustrate its application to a
nonideal prepare-and-measure BB84 implementation, in which Alice's source
emits four arbitrary but characterised pure states and Bob performs
uncharacterised measurements, for which a suitable fidelity bound is already
given in \cite{ref:w2013}. In this setting, Alice's source emits two states,
which we call the ``$x$-basis'' states and denote $\ket{\beta}$ and
$\ket{\beta'}$, and Bob performs an ``$x$-basis'' measurement, in addition to
the $z$-basis states $\ket{\alpha}$ and $\ket{\alpha'}$ and $z$-basis
measurement previously described. (We adopt this nomenclature even if the
source states and measurements do not satisfy the ideal BB84 relations.) We
assume Alice's source is characterised by an angular parameter $\theta$,
defined in terms of the source states by
\begin{equation}
  \label{eq:theta_def}
  \sqrt{1 + \abs{\sin(\theta)}}
  = \tfrac{1}{2} \babs{\braket{\alpha}{\beta} +
    \braket{\alpha'}{\beta} + \braket{\alpha}{\beta'} -
    \braket{\alpha'}{\beta'}} \,,
\end{equation}
wherever the right-hand side of \eqref{eq:theta_def} is greater than 1. Given
this characterisation, the fidelity between Eve's marginals of the $z$ states
is lower bounded by
\begin{IEEEeqnarray}{rCl}
  \label{eq:fidel_theta}
  F(\rho\0_{\rE}, \rho\1_{\rE})
  &\geq& \abs{\sin(\theta)} D(\sigma\0_{\rB}, \sigma\1_{\rB})
  \IEEEnonumber \\
  &&-\> \abs{\cos(\theta)}
  \sqrt{1 - D(\sigma\0_{\rB}, \sigma\1_{\rB})^{2}} \,,
\end{IEEEeqnarray}
where $\sigma\0_{\rB} = \Tr_{\rE}[\proj{\beta}]$, $\sigma\1_{\rB} =
\Tr_{\\rE}[\proj{\beta'}]$, and $D(\sigma\0_{\rB}, \sigma\1_{\rB}) =
\tfrac{1}{2} \trnorm{\sigma\0_{\rB} - \sigma\1_{\rB}}$ is the trace distance
between Bob's marginals of the $x$ states \cite{ref:w2013}. The trace
distance itself is lower bounded by $D(\sigma\0_{\rB}, \sigma\1_{\rB}) \geq
\abs{1 - 2 \delta_{x}}$ in terms of the $x$-basis error rate, regardless of
how Bob's measurement apparatus is oriented. Explicitly combining these with
the generic key-rate expression \eqref{eq:r_fidel_bound}, we obtain the bound
\begin{IEEEeqnarray}{rCl}
  \label{eq:r_theta_noise}
  r &\geq& 1 + h \Bigro{\tfrac{1}{2}
    + \tfrac{1}{2} \sqrt{1 - 4 \eta (1 - \eta)
      \bigro{1 - f_{\theta}(\abs{1 - 2 \delta_{x}})^{2}}}} \IEEEnonumber \\
  &&-\> h \bigro{\tfrac{1}{2}
    + \tfrac{1}{2} f_{\theta}(\abs{1 - 2 \delta_{x}})}
  - h(\tilde{\delta}_{z}) \,,
\end{IEEEeqnarray}
with $\tilde{\delta}_{z} = (1 - \eta) \delta_{z} + \eta (1 - \delta_{z})$ and
$f_{\theta}$ defined by
\begin{equation}
  \label{eq:def_f_theta}
  f_{\theta}(x)
  = \begin{cases}
    \abs{\sin(\theta)} x - \abs{\cos(\theta)} \sqrt{1 - x^{2}}
    &: x \geq \abs{\cos(\theta)} \\
    0 &: x \leq \abs{\cos(\theta)}
  \end{cases} \,,
\end{equation}
for the key rate with local randomisation applied. For $\eta = 0$, we recover
the key rate
\begin{equation}
  \label{eq:r_theta_clean}
  r \geq 1 - h \bigro{\tfrac{1}{2}
    + \tfrac{1}{2} f_{\theta}(\abs{1 - 2 \delta_{x}})}
  - h(\delta_{z})
\end{equation}
given in \cite{ref:w2013}, which itself coincides with the rate derived in
\cite{ref:mls2010} in the setting under consideration here.

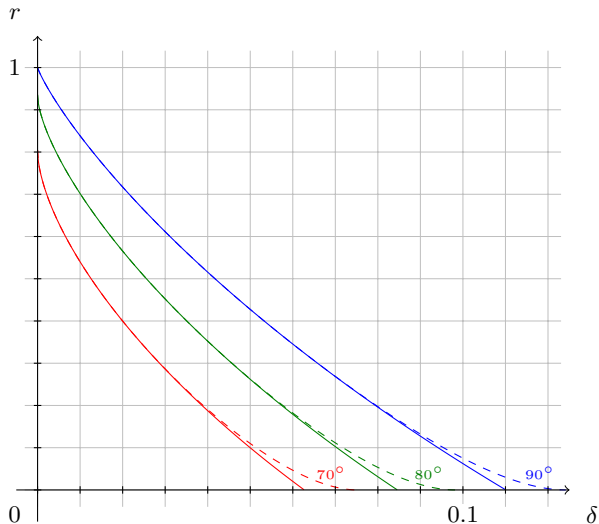
\begin{figure}[htbp]
  \centering
  \begin{tikzpicture}[xscale=80,yscale=8,inner sep=1pt]
    \draw[gridlines,xstep=0.01,ystep=0.1] (-0.003,-0.04) grid (0.123,1.04);

    \draw[smooth,color=blue] plot file{plots/r_90_clean.table};
    \draw[smooth,color=blue,dashed] plot file{plots/r_90_noise.table}
      node[above left=3pt] {\tiny $90^{\circ}$};

    \draw[smooth,color=darkgreen] plot file{plots/r_80_clean.table};
    \draw[smooth,color=darkgreen,dashed] plot file{plots/r_80_noise.table}
      node[above left=3pt] {\tiny $80^{\circ}$};

    \draw[smooth,color=red] plot file{plots/r_70_clean.table};
    \draw[smooth,color=red,dashed] plot file{plots/r_70_noise.table}
      node[above left=3pt] {\tiny $70^{\circ}$};

    \draw[->] (-0.005,0) -- (0.125,0)
      node[below right=5pt,fill=white]{$\delta$};
    \draw[->] (0,-0.075) -- (0,1.075)
      node[above left=5pt,fill=white]{$r$};

    \foreach \x in {0,0.01,...,0.1201}
      \draw (\x, -0.25pt) -- (\x, 0.25pt);

    \foreach \x in {0,0.1,...,1.01}
      \draw (-0.025pt, \x) -- (0.025pt, \x);

    \draw (0,0) node[below left=5pt,fill=white]{$0$};
    \draw (0.1,0) node [below=5pt,fill=white]{$0.1$};
    \draw (0,1) node[left=5pt,fill=white]{$1$};
  \end{tikzpicture}
  \caption{Key rates for $\theta = 90^{\circ}$, $\theta = 80^{\circ}$, and
    $\theta = 70^{\circ}$ with (dashed curves) and without (solid curves)
    local randomisation, for $\delta = \delta_{z} = \delta_{x}$.}
  \label{fig:keyrates_pm}
\end{figure}

The rates \eqref{eq:r_theta_noise} and \eqref{eq:r_theta_clean} (with and
without local randomisation, respectively) are illustrated for a few values
of $\theta$ in Fig.~\ref{fig:keyrates_pm}, assuming symmetric errors (i.e.,
$\delta_{z} = \delta_{x} = \delta$) for simplicity. The depicted rates with
local randomisation were found by numerically maximising
\eqref{eq:r_theta_noise} over $\eta$. For $\theta = \pi/2 = 90^{\circ}$,
corresponding to an ideal BB84 source, we recover the Shor-Preskill rate
\cite{ref:sp2000} and the improvement with local randomisation depicted in
Fig.~2 of Ref.~\cite{ref:rgk2005}.

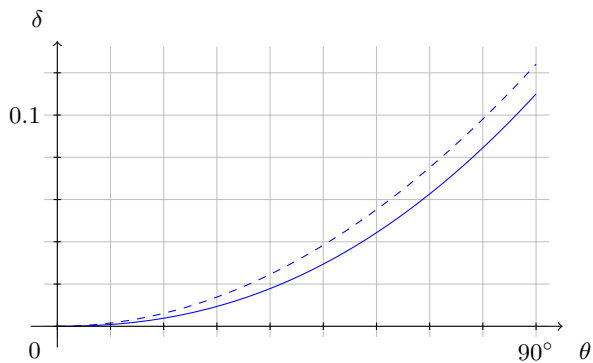
\begin{figure}[htbp]
  \centering
  \begin{tikzpicture}[xscale=0.1,yscale=40,inner sep=1pt]
    \draw[gridlines,xstep=10,ystep=0.02] (-2.5,-0.005) grid (92.5,0.1325);

    \draw[smooth,color=blue,dashed] plot file{plots/pm_thresh_noise.table};
    \draw[smooth,color=blue] plot file{plots/pm_thresh_clean.table};

    \draw[->] (-5,0) -- (95,0)
      node[below right=5pt,fill=white]{$\theta$};
    \draw[->] (0,-0.01) -- (0,0.135)
      node[above left=4pt,fill=white]{$\delta$};

    \foreach \x in {0,10,...,90.1}
      \draw (\x, -0.05pt) -- (\x, 0.05pt);

    \foreach \x in {0,0.02,...,0.121}
      \draw (-20pt, \x) -- (20pt, \x);

    \draw (0,0) node[below left=5pt,fill=white]{$0$};
    \draw (90,0) node [below=5pt,fill=white]{$90^{\circ}$};
    \draw (0,0.1) node[left=5pt,fill=white]{$0.1$};
  \end{tikzpicture}
  \caption{Threshold error rate with (dashed curve) and without (solid curve)
    local randomisation, for $0 \leq \theta \leq 90^{\circ}$.}
  \label{fig:pm_thresh}
\end{figure}

The threshold error rates, i.e., the error rates for which the key rates
\eqref{eq:r_theta_clean} without preprocessing and \eqref{eq:r_theta_noise}
with optimal local randomisation become zero, again for $\delta_{z} =
\delta_{x} = \delta$, are depicted in Fig.~\ref{fig:pm_thresh} as a function
of $\theta$. The threshold curve with local randomisation was found by
identifying the corresponding root of \eqref{eq:thresh_root}. For $\theta =
\pi/2 = 90^{\circ}$ we recover the threshold error rates of $\delta \approx
12.4120\%$ and $\delta \approx 11.0028\%$ originally found in
Refs.~\cite{ref:kgr2005} and \cite{ref:sp2000}, respectively. For an ideal
BB84 implementation, this corresponds to a relative increase of around
12.81\% to the provably tolerable channel noise. This difference becomes more
significant as $\theta$ decreases: for instance the relative improvement
becomes around 20.00\% ($\delta \approx 7.5191\%$ compared with $\delta
\approx 6.2660\%$) for $\theta = 70^{\circ}$, around 33.84\% ($\delta \approx
3.1120\%$ vs $\delta \approx 2.3251\%$) for $\theta = 45^{\circ}$, and around
83.38\% ($\delta \approx 0.1538\%$ vs $\delta \approx 0.08390\%$) if $\theta$
is as low as $10^{\circ}$, indicating that the benefit of additional
preprocessing becomes more pronounced for a realistic BB84 implementation
expected to suffer from device imprecisions.

\section{Optimality of key-rate bound\label{sec:optimality}}

For the nonideal BB84 implementation considered in the preceding section, the
key-rate bound \eqref{eq:r_theta_noise} is tight in the sense that the
Devetak-Winter rate \eqref{eq:dw_rate} can be attained for all values of the
independent variables $\theta$, $\delta_{z}$, $\delta_{x}$, and $\eta$. This
is demonstrated here by the explicit construction of a family of source
states and optimal unitary attacks. Equality between the right-hand sides of
\eqref{eq:dw_rate} and \eqref{eq:r_theta_noise} requires the conditional von
Neumann entropy bound \eqref{eq:H_ZE_noise} and the fidelity bound
\eqref{eq:fidel_theta} to hold with equality simultaneously, which helps in
the determination of an optimal attack. First, note that in the case of an
equality, \eqref{eq:fidel_theta} rearranges to
\begin{equation}
  \label{eq:theta_FZ_DX}
  \abs{\sin(\theta)} = F_{Z} D_{X}
  + \sqrt{1 - F\subsup{Z}{2}} \sqrt{1 - D\subsup{X}{2}} \,,
\end{equation}
with $F_{Z} = F(\rho\0_{\rE}, \rho\1_{\rE})$ and $D_{X} = D(\sigma\0_{\rB},
\sigma\1_{\rB})$ and the condition $F_{Z} \leq
D_{X}$. Equation~\eqref{eq:theta_FZ_DX} can equivalently be reexpressed as
\begin{IEEEeqnarray}{rl}
  \label{eq:theta_FZ_DX_root}
  \sqrt{1 + \abs{\sin(\theta)}} = \frac{1}{\sqrt{2}} \Bigro{
    &\sqrt{1 + F_{Z}} \sqrt{1 + D_{X}} \IEEEnonumber \\
    &+\> \sqrt{1 - F_{Z}} \sqrt{1 - D_{X}}} \,.
\end{IEEEeqnarray}
Consequently, our goal will be to construct source states such that the
definition of the source characterisation \eqref{eq:theta_def} equals the
right-hand side of \eqref{eq:theta_FZ_DX_root}.

Requiring $D(\rho\0_{\rB}, \rho\1_{\rB}) = D_{Z}$ suggests setting the
$z$-basis states to the form
\begin{IEEEeqnarray}{rCl}
  \label{eq:alpha0_pm_def}
  \ket{\alpha}
  &=& \sqrt{\tfrac{1 + D_{Z}}{2}} \ket{0}_{\rB} \ket{\psi_{0}}_{\rE}
  + \sqrt{\tfrac{1 - D_{Z}}{2}} \ket{1}_{\rB} \ket{\psi'_{1}}_{\rE} \,, \\
  \label{eq:alpha1_pm_def}
  \ket{\alpha'}
  &=& \sqrt{\tfrac{1 - D_{Z}}{2}} \ket{0}_{\rB} \ket{\psi_{1}}_{\rE}
  + \sqrt{\tfrac{1 + D_{Z}}{2}} \ket{1}_{\rB} \ket{\psi'_{0}}_{\rE} \,,
\end{IEEEeqnarray}
with $\ket{0}_{\rB}$ and $\ket{1}_{\rB}$ orthonormal. The trace distance
$D(\rho\0_{\rB}, \rho\1_{\rB})$ will equal $D_{Z}$ if
$\braket{\psi_{0}}{\psi'_{1}} = \braket{\psi_{1}}{\psi'_{0}} = 0$. In order
for the fidelity $F(\rho\0_{\rE}, \rho\1_{\rE})$ to equal $F_{Z}$, and in
such a way that the von Neumann entropy bound \eqref{eq:H_ZE_noise} becomes
an equality, we additionally require $\braket{\psi_{0}}{\psi_{1}} =
\braket{\psi'_{0}}{\psi'_{1}} = 0$ and $\braket{\psi_{0}}{\psi'_{0}} =
\braket{\psi_{1}}{\psi'_{1}} = F_{Z} \in \mathbb{R}^{+}$, such that
$\{\ket{\psi_{0}}, \ket{\psi'_{0}}\}$ and $\{\ket{\psi_{1}},
\ket{\psi'_{1}}\}$ span two mutually orthogonal subspaces. Note that, with
these definitions, $\ket{\alpha}$ and $\ket{\alpha'}$ are normalised and
orthogonal.

The right-hand side of the source characterisation \eqref{eq:theta_def} can
be reexpressed as $\frac{1}{\sqrt{2}} \babs{\braket{\alpha_{+}}{\beta} +
  \braket{\alpha_{-}}{\beta'}}$ with $\ket{\alpha_{\pm}} = \frac{1}{\sqrt{2}}
\bigro{\ket{\alpha} \pm \ket{\alpha'}}$. Introducing, for convenience, the
states
\begin{IEEEeqnarray}{rCl+rCl}
  \ket{\alpha_{k}} &=& \ket{0}_{\rB} \ket{\psi_{k}}_{\rE} \,, &
  \ket{\alpha'_{k}} &=& \ket{1}_{\rB} \ket{\psi'_{k}}_{\rE} \,,
\end{IEEEeqnarray}
and $\ket{\alpha_{k}^{\pm}} = \frac{1}{\sqrt{2}} \bigro{\ket{\alpha_{k}} \pm
  \ket{\alpha'_{k}}}$, $k \in \{0, 1\}$, we find
\begin{IEEEeqnarray}{rCl}
  \ket{\alpha_{k}^{+}}
  &=& \sqrt{\tfrac{1 + F_{Z}}{2}} \ket{+}_{\rB} \ket{\psi_{k}^{+}}_{\rE}
  + \sqrt{\tfrac{1 - F_{Z}}{2}} \ket{-}_{\rB} \ket{\psi_{k}^{-}}_{\rE} \,, \\
  \ket{\alpha_{k}^{-}}
  &=& \sqrt{\tfrac{1 - F_{Z}}{2}} \ket{+}_{\rB} \ket{\psi_{k}^{-}}_{\rE}
  + \sqrt{\tfrac{1 + F_{Z}}{2}} \ket{-}_{\rB} \ket{\psi_{k}^{+}}_{\rE} \,,
\end{IEEEeqnarray}
where $\ket{\pm}_{\rB} = \frac{1}{\sqrt{2}} \bigro{\ket{0}_{\rB} +
  \ket{1}_{\rB}}$ and the states
\begin{equation}
  \ket{\psi_{k}^{\pm}}_{\rE}
  = \frac{\ket{\psi_{k}}_{\rE} \pm \ket{\psi'_{k}}_{\rE}}{
    \sqrt{2 \pm 2 F_{Z}}}
\end{equation}
are orthonormal. In terms of $\ket{\alpha_{k}^{\pm}}$,
\begin{IEEEeqnarray}{rCl}
  \ket{\alpha_{+}} &=& \sqrt{\tfrac{1 + D_{Z}}{2}} \ket{\alpha_{0}^{+}}
  + \sqrt{\tfrac{1 - D_{Z}}{2}} \ket{\alpha_{1}^{+}} \,, \\
  \ket{\alpha_{-}} &=& \sqrt{\tfrac{1 + D_{Z}}{2}} \ket{\alpha_{0}^{-}}
  - \sqrt{\tfrac{1 - D_{Z}}{2}} \ket{\alpha_{1}^{-}} \,.
\end{IEEEeqnarray}
It is then fairly straightforward to construct $x$-basis states for which the
right-hand side of the source characterisation \eqref{eq:theta_def} will take
the form of the right-hand side of \eqref{eq:theta_FZ_DX_root}. We set
\begin{IEEEeqnarray}{rCl}
  \ket{\beta_{k}}
  &=& \sqrt{\tfrac{1 + D_{X}}{2}} \ket{+}_{\rB} \ket{\psi_{k}^{+}}_{\rE}
  + \sqrt{\tfrac{1 - D_{X}}{2}} \ket{-}_{\rB} \ket{\psi_{k}^{-}}_{\rE} \,, \\
  \ket{\beta'_{k}}
  &=& \sqrt{\tfrac{1 - D_{X}}{2}} \ket{+}_{\rB} \ket{\psi_{k}^{-}}_{\rE}
  + \sqrt{\tfrac{1 + D_{X}}{2}} \ket{-}_{\rB} \ket{\psi_{k}^{+}}_{\rE} \,,
\end{IEEEeqnarray}
and
\begin{IEEEeqnarray}{rCl}
  \label{eq:beta0_pm_def}
  \ket{\beta} &=& \sqrt{\tfrac{1 + D_{Z}}{2}} \ket{\beta_{0}}
  + \sqrt{\tfrac{1 - D_{Z}}{2}} \ket{\beta_{1}} \,, \\
  \label{eq:beta1_pm_def}
  \ket{\beta'} &=& \sqrt{\tfrac{1 + D_{Z}}{2}} \ket{\beta'_{0}} -
  \sqrt{\tfrac{1 - D_{Z}}{2}} \ket{\beta'_{1}} \,.
\end{IEEEeqnarray}
With these definitions we find
\begin{IEEEeqnarray}{rl}
  \braket{\alpha_{+}}{\beta} = \braket{\alpha_{-}}{\beta'}
  = \frac{1}{2} \Bigro{& \sqrt{1 + F_{Z}} \sqrt{1 + D_{X}} \IEEEnonumber \\
    &+\> \sqrt{1 - F_{Z}} \sqrt{1 - D_{X}}} \,,
\end{IEEEeqnarray}
independently of $D_{Z}$, from which we recover the right-hand side of the
rearrangement \eqref{eq:theta_FZ_DX_root} of the fidelity bound
\eqref{eq:fidel_theta}.

Explicitly, from the expressions \eqref{eq:alpha0_pm_def},
\eqref{eq:alpha1_pm_def}, \eqref{eq:beta0_pm_def}, and
\eqref{eq:beta1_pm_def} for the $z$- and $x$-basis states, Bob's marginals
are given by
\begin{IEEEeqnarray}{rCl}
  \rho\0_{\rB} &=& \tfrac{1 + D_{Z}}{2} \proj{0}_{\rB}
  + \tfrac{1 - D_{Z}}{2} \proj{1}_{\rB} \,, \\
  \rho\1_{\rB} &=& \tfrac{1 - D_{Z}}{2} \proj{0}_{\rB}
  + \tfrac{1 + D_{Z}}{2} \proj{1}_{\rB} \,, \\
  \sigma\0_{\rB} &=& \tfrac{1 + D_{X}}{2} \proj{+}_{\rB}
  + \tfrac{1 - D_{X}}{2} \proj{-}_{\rB} \,, \\
  \sigma\1_{\rB} &=& \tfrac{1 - D_{X}}{2} \proj{+}_{\rB}
  + \tfrac{1 + D_{X}}{2} \proj{-}_{\rB} \,.
\end{IEEEeqnarray}
Consequently, Alice and Bob detect errors at the rates $\delta_{z} =
\tfrac{1}{2} - \tfrac{1}{2} D_{Z}$ and $\delta_{x} = \tfrac{1}{2} -
\tfrac{1}{2} D_{X}$ if Bob measures (optimally) in the $\sz$ and $\sx$
bases. Likewise, Eve's marginals of the $z$ states are given by
\begin{IEEEeqnarray}{rCl}
  \rho\0_{\rE} &=& \tfrac{1 + D_{Z}}{2} \proj{\psi_{0}}_{\rE}
  + \tfrac{1 - D_{Z}}{2} \proj{\psi'_{1}}_{\rE} \,, \\
  \rho\1_{\rE} &=& \tfrac{1 - D_{Z}}{2} \proj{\psi_{1}}_{\rE}
  + \tfrac{1 + D_{Z}}{2} \proj{\psi'_{0}}_{\rE} \,,
\end{IEEEeqnarray}
for which one can readily verify that $F(\rho\0_{\rE}, \rho\1_{\rE}) =
\trnorm{\sqrt{\rho\0_{\rE}} \sqrt{\rho\1_{\rE}}} = F_{Z}$ and, for any $p, q
\geq 0$ and $p + q = 1$,
\begin{IEEEeqnarray}{rCl}
  \label{eq:S_rhos}
  S\bigro{p \rho\0_{\rE} + q \rho\1_{\rE}}
  &=& h\bigro{\tfrac{1}{2} + \tfrac{1}{2} D_{Z}} \IEEEnonumber \\
  &&+\> \tfrac{1 + D_{Z}}{2}
  S\bigro{p \proj{\psi_{0}}_{\rE} + q \proj{\psi'_{0}}_{\rE}}
  \IEEEnonumber \\
  &&+\> \tfrac{1 - D_{Z}}{2}
  S\bigro{p \proj{\psi'_{1}}_{\rE} + q \proj{\psi_{1}}_{\rE}}
  \IEEEnonumber \\
  &=& h\bigro{\tfrac{1}{2} + \tfrac{1}{2} D_{Z}} \IEEEnonumber \\
  &&+\> h \Bigro{\tfrac{1}{2}
    + \tfrac{1}{2} \sqrt{1 - 4pq (1 - F\subsup{Z}{2})}} \,.
\end{IEEEeqnarray}
Using \eqref{eq:S_rhos} to directly evaluate the expression
\eqref{eq:H_ZE_rhos} for the conditional von Neumann entropy $\Hq(Z \mid
\rE)$, we find that its bound \eqref{eq:H_ZE_noise} in terms of fidelity
$F(\rho\0_{\rE}, \rho\1_{\rE})$ is attained with equality for the entire
family of sources and attacks just constructed.

Equations~\eqref{eq:alpha0_pm_def}, \eqref{eq:alpha1_pm_def},
\eqref{eq:beta0_pm_def}, and \eqref{eq:beta1_pm_def} give the optimal attack
for a family of sources identified by the relations
\begin{equation}
  \braket{\alpha}{\alpha'} = \braket{\beta}{\beta'} = 0
\end{equation}
and
\begin{equation}
  \braket{\alpha}{\beta} = \braket{\alpha'}{\beta}
  = \braket{\alpha}{\beta'} = - \braket{\alpha'}{\beta'} 
  = \tfrac{\sqrt{1 + \abs{\sin(\theta)}}}{2} \,,
\end{equation}
for which the bound \eqref{eq:r_theta_noise} on the Devetak-Winter rate is
attained with equality independently of the fraction $\eta$ of bits flipped
by Alice in the local randomisation preprocessing step. The family of optimal
attacks given here generalises the optimal individual attack derived for an
ideal BB84 source in \cite{ref:fg1997}, which is recovered for
$\abs{\sin(\theta)} = 1$ or, equivalently, by setting $F_{Z} =
D_{X}$. Another extreme worth noting is the case $F_{Z} = 0$ and $D_{Z} =
D_{X} = 1$, in which case $\abs{\sin(\theta)} = 0$ and
\begin{IEEEeqnarray}{rCl+rCl}
  \ket{\alpha} &=& \ket{0}_{\rB} \ket{0}_{\rE} \,, &
  \ket{\alpha'} &=& \ket{1}_{\rB} \ket{1}_{\rE} \,, \\
  \ket{\beta} &=& \ket{+}_{\rB} \ket{+}_{\rE} \,, &
  \ket{\beta'} &=& \ket{-}_{\rB} \ket{+}_{\rE} \,,
\end{IEEEeqnarray}
i.e., the adversary acquires perfect copies of Bob's $z$-basis states without
introducing any errors.

\section{Conclusion}

This \work{} described how the local randomisation preprocessing technique
proposed by Kraus \etal{} in \cite{ref:kgr2005,ref:rgk2005} can be
incorporated into the security analysis introduced in \cite{ref:w2013} for
the BB84 protocol. The improvement to the key rate and tolerable channel
noise was explicitly quantified for an imperfect BB84 implementation in which
Alice's source emits four arbitrary but characterised pure states and Bob's
measurements are left largely uncharacterised. The improvement becomes more
significant if the source is imperfect. The asymptotic key-rate bound
\eqref{eq:r_theta_noise} was shown to be tight given the source
characterisation parameter $\theta$ defined in \eqref{eq:theta_def}, and is
attained for the family of source states and optimal collective attacks
constructed in Sec.~\ref{sec:optimality} if Bob performs the ideal $\sz$ and
$\sx$ measurements.

The setting described in Sec.~\ref{sec:scenario} was left somewhat generic as
the method is not necessarily limited to just the BB84 protocol itself. In
particular it has already been found to apply to a semi-device-independent
QKD protocol in which Alice's source and Bob's measurements are assumed two
dimensional \cite{ref:wpYYYY}. The device-independent protocol studied in
\cite{ref:ab2007} may also be a candidate; this is suggested by the fact that
the conditional von Neumann entropy bound \eqref{eq:H_ZE_noise} is attained
with equality for the optimal collective attack derived in \cite{ref:ab2007}.

\begin{acknowledgments}
  S.~Pironio offered helpful comments on an early draft of this \work{}. This
  work was supported by the EU projects Q-Essence and QAlgo, the CHIST-ERA
  DIQIP project, the Interuniversity Attraction Poles Photonics@be Programme
  (Belgian Science Policy), and the FRS-FNRS under project DIQIP.\@ The
  author is supported by a Belgian Fonds pour la Formation \`a{} la Recherche
  dans l'Industrie et dans l'Agriculture (F.R.I.A.) doctoral grant.
\end{acknowledgments}

\appendix

\section{Equivalent entanglement-based protocol\label{sec:virtual_ent}}

The main problem addressed in this \work{} is the derivation of key rates for
variants of the prepare-and-measure BB84 protocol secure against an adversary
limited to collective attacks. In the case of entanglement-based QKD,
security against collective attacks is known to imply security against
general attacks in the asymptotic limit, at least under the assumption of a
dimension bound, the assumption that Alice's and Bob's measurements are
memoryless, and if a symmetrisation procedure is applied
\cite{ref:ckr2009}. In this section, we briefly review how the BB84 protocol
considered in Sec.~\ref{sec:pmbb84} could be recast in the form of an
equivalent entanglement-based protocol.

The starting point, already considered by the authors of
\cite{ref:gl2004,ref:k2009,ref:mls2010}, is that Alice could prepare the four
source states by equivalently preparing and distributing an entangled
``coin'' state of the form
\begin{IEEEeqnarray}{rCl}
  \label{eq:coin_state}
  \ket{\Psi}_{\mathrm{ABE}} &=& c_{00} \ket{00}_{\rA} \ket{\alpha}_{\rBE}
  + c_{01} \ket{01}_{\rA} \ket{\alpha'}_{\rBE} \IEEEnonumber \\
  &&+\> c_{10} \ket{10}_{\rA} \ket{\beta}_{\rBE}
  + c_{11} \ket{11}_{\rA} \ket{\beta'}_{\rBE}
\end{IEEEeqnarray}
and determining her bit and basis choice by measuring in the (orthonormal)
$\{\ket{00}_{\rA}, \ket{01}_{\rA}, \ket{10}_{\rA}, \ket{11}_{\rA}\}$
basis. In \eqref{eq:coin_state}, the coefficients $c_{ij}, i, j \in \{0, 1\}$
determine the probability of Alice selecting a particular bit and basis and
can always be taken to be real and nonnegative. In the setting considered in
Sec.~\ref{sec:pmbb84}, Alice uses the $z$ basis the majority of the time and
chooses between the two states in each basis equiprobably. For the amplitudes
$c_{ij}$, this translates to
\begin{IEEEeqnarray}{rCcCl}
  c_{00} &=& c_{01} &=& \sqrt{\frac{1 - \varepsilon}{2}} \,, \\
  c_{10} &=& c_{11} &=& \sqrt{\frac{\varepsilon}{2}}
\end{IEEEeqnarray}
for some given $\varepsilon$ close to zero. The states $\ket{\alpha},
\ket{\alpha'}, \ket{\beta}, \ket{\beta'} \in \hilb_{\rB} \otimes \hilb_{\rE}$
shared by Bob and Eve correspond to the $z$- and $x$-basis states and the
relations between them are fully specified, such that \eqref{eq:coin_state}
is determined up to an overall unitary on $\hilb_{\rB} \otimes \hilb_{\rE}$.

The virtual protocol described so far is not a secure entanglement-based
protocol in the usual sense, as it requires Alice and Bob to trust that the
shared entangled state is of the form given in \eqref{eq:coin_state}. To
remove this, note that it is equivalent for Alice to know the marginal
density operator $\rho_{\rA} = \Tr_{\rBE}\bigsq{\proj{\Psi}_{\rABE}}$, as all
purifications of $\rho_{\rA}$ are related by unitaries on $\hilb_{\rB}
\otimes \hilb_{\rE}$. Specifically, if $\ket{\Psi}_{\rABE}$ is any
purification of $\rho_{\rA}$, the $z$ and $x$ states and amplitutes appearing
in \eqref{eq:coin_state} can be recovered by
\begin{IEEEeqnarray}{rCl}
  c_{00} \ket{\alpha}_{\rBE}
  &=& (\bra{00}_{\rA} \otimes \id_{\rBE}) \ket{\Psi}_{\rABE} \,, \\
  c_{01} \ket{\alpha'}_{\rBE}
  &=& (\bra{01}_{\rA} \otimes \id_{\rBE}) \ket{\Psi}_{\rABE} \,, \\
  c_{10} \ket{\beta}_{\rBE}
  &=& (\bra{10}_{\rA} \otimes \id_{\rBE}) \ket{\Psi}_{\rABE} \,, \\
  c_{11} \ket{\beta'}_{\rBE}
  &=& (\bra{11}_{\rA} \otimes \id_{\rBE}) \ket{\Psi}_{\rABE} \,.
\end{IEEEeqnarray}
Since $\rho_{\rA}$ is reconstructable by tomography, Alice no longer needs to
trust that the entangled state is of the form \eqref{eq:coin_state}, provided
that she performs additional tomographic measurements to determine or verify
that $\rho_{\rA}$ is of a particular desired form as part of the virtual
entanglement-based protocol outlined here.

\section{Convexity of conditional entropy bound\label{sec:convexity}}

The right-hand side of \eqref{eq:H_ZE_noise} has the form
\begin{equation}
  \label{eq:H_phi_F}
  H(F) = 1 + \phi(R) - \phi(F) \,,
\end{equation}
where, for convenience, we have set
\begin{equation}
  R = \sqrt{\lambda + \mu F^{2}} \,,
\end{equation}
$\lambda = (1 - 2 \eta)^{2}$ and $\mu = 4 \eta (1 - \eta)$ (such that $0 \leq
\lambda, \mu \leq 1$ and $\lambda + \mu = 1$), and the function $\phi$ is
defined by
\begin{IEEEeqnarray}{rCl}
  \phi(x) &=& h(\tfrac{1}{2} + \tfrac{1}{2} x) \IEEEnonumber \\
  &=& 1 - \tfrac{1}{2} (1 + x) \log(1 + x) \IEEEnonumber \\
  &&-\> \tfrac{1}{2} (1 - x) \log(1 - x)
\end{IEEEeqnarray}
for $-1 < x < 1$ and $\phi(1) = \phi(-1) = 0$.

If $\mu = 1$ (and $\lambda = 0$), \eqref{eq:H_phi_F} reduces to $H(F) =
1$. In the following we show that, for $\mu < 1$, $H$ is a convex function by
showing that its second derivative in $F$ is nonnegative. Since its global
minimum is $H(0) = 0$, it will follow that $H$ is an increasing function over
the range $0 \leq F \leq 1$.

We first evaluate the first and second derivatives of $\phi$; respectively,
they are
\begin{equation}
  \phi'(x) = - \tfrac{1}{2} \log \Bigro{\frac{1 + x}{1 - x}}
\end{equation}
and
\begin{equation}
  \phi''(x) = - \frac{1}{\ln(2)} \frac{1}{1 - x^{2}} \,.
\end{equation}
For the first and second derivatives of $R$ (viewed as a function of $F$), we
obtain $R' = \mu F / R$ and $R'' = \lambda \mu / R^{3}$. In terms of $\phi$
and its derivatives and $R$, the first and second derivatives of $H$ are
\begin{equation}
  \label{eq:H_1st_deriv}
  H'(F) = \phi'(R) \frac{\mu F}{R} - \phi'(F) \,,
\end{equation}
and
\begin{equation}
  \label{eq:H_2nd_deriv}
  H''(F) = \phi''(R) \frac{\mu^{2} F^{2}}{R^{2}}
  + \phi'(R) \frac{\lambda \mu}{R^{3}} - \phi''(F) \,.
\end{equation}
Using that $\phi''(F) = \mu \phi''(R)$ and that $\mu F^{2} - R^{2} = \mu - 1
= - \lambda$, \eqref{eq:H_2nd_deriv} can be rearranged to
\begin{IEEEeqnarray}{rCl}
  H''(F) &=& \frac{\lambda \mu}{R^{3}} \Bigro{- R \phi''(R) + \phi'(R)}
  \IEEEnonumber \\
  &=& \frac{1}{\ln(2)} \frac{\lambda \mu}{R^{3}}
  \biggsq{\frac{R}{1 - R^{2}}
    - \tfrac{1}{2} \ln \biggro{\frac{1 + R}{1 - R}}} \IEEEnonumber \\
  &=& \frac{1}{4 \ln(2)} \frac{\lambda \mu}{R^{3}}
  \Bigro{Z - \frac{1}{Z} - 2 \ln(Z)} \,,
\end{IEEEeqnarray}
where we set $Z = (1 + R) / (1 - R)$ and we used that
\begin{equation}
  \frac{4 R}{1 - R^{2}}
  = \frac{(1 + R)^{2} - (1 - R)^{2}}{(1 + R)(1 - R)}
  = Z - \frac{1}{Z} \,.
\end{equation}
Finally, we note that $(\lambda \mu) / (4 \ln(2) R^{3}) \geq 0$ and that, for
$Z \geq 1$,
\begin{IEEEeqnarray}{rCl}
  Z - \frac{1}{Z} - 2 \ln(Z)
  &=& \int_{1}^{Z} \dd z \Bigro{1 + \frac{1}{z^{2}}}
  - 2 \int_{1}^{Z} \dd z \frac{1}{z} \IEEEnonumber \\
  &=& \int_{1}^{Z} \dd z \Bigro{1 - \frac{1}{z}}^{2} \IEEEnonumber \\
  &\geq& 0 \,,
\end{IEEEeqnarray}
which together imply $H''(F) \geq 0$.

From \eqref{eq:H_1st_deriv}, we see that $H'(0) = 0$, confirming that $F = 0$
is at least a local extremum. Since $H$ is convex, the only possibility is
that $F = 0$ is, in fact, the global minimum, in turn implying that $H$ is an
increasing function of $F$ over the range $0 \leq F \leq 1$.

\bibliography{qkd_noise}

\end{document}